# System calibration method for Fourier ptychographic microscopy

**An Pan,**[1,2] **Yan Zhang,**[1,2] **Tianyu Zhao,**[1,2] **Zhaojun Wang,**[1,2] **Dan Dan**[1,2] **and Baoli Yao,**[1,*]

[1]*State Key Laboratory of Transient Optics and Photonics, Xi'an Institute of Optics and Precision Mechanics, Chinese Academy of Sciences, Xi'an 710119, China*
[2]*University of Chinese Academy of Sciences, Beijing 100049, China*
*\*yaobl@opt.ac.cn*

**Abstract:** Fourier ptychographic microscopy (FPM) is a recently proposed quantitative phase imaging technique with high resolution and wide field-of-view (FOV). In current FPM imaging platforms, systematic error sources come from the aberrations, LED intensity fluctuation, parameter imperfections and noise, which will severely corrupt the reconstruction results with artifacts. Although these problems have been researched and some special methods have been proposed respectively, there is no method to solve all of them. However, the systematic error is a mixture of various sources in the real situation. It is difficult to distinguish a kind of error source from another due to the similar artifacts. To this end, we report a system calibration procedure, termed SC-FPM, based on the simulated annealing (SA) algorithm, LED intensity correction and adaptive step-size strategy, which involves the evaluation of an error matric at each iteration step, followed by the re-estimation of accurate parameters. The great performance has been achieved both in simulation and experiments. The reported system calibration scheme improves the robustness of FPM and relaxes the experiment conditions, which makes the FPM more pragmatic.

## 1. Introduction

Fourier ptychographic microscopy (FPM) [1-3] is a recently proposed quantitative phase imaging technique with high resolution and wide field-of-view (FOV). By recording multiple low-resolution (LR) intensity images of the sample from angle-varied illumination and iteratively stitching these different LR intensity images together in the Fourier space, FPM recovers a high-resolution (HR) complex amplitude image of the sample with a large FOV, which overcomes the physical space-bandwidth-product (SBP) limit of a low numerical aperture (NA) imaging system. Instead of making a compromise between a large FOV and HR, FPM achieves both of them by trading acquisition speed, which shares its root with conventional ptychography [4-10] and synthetic aperture imaging [11-14]. The final reconstruction resolution is determined by the sum of the objective lens and illumination NAs [15]. Due to its flexible setup, perfect performance and rich redundancy of the acquired data, FPM has been widely applied in 3D imaging [16, 17], fluorescence imaging [18, 19], multiplexing imaging [20, 21], high-speed imaging [22, 23].

In current FPM imaging platforms, systematic error sources mainly come from aberrations, LED intensity fluctuation, parameter imperfections and noise, which will severely degrade the reconstruction results with artifacts. Thus a series of studies have been reported to relax or overcome them respectively [24-33]. Ou et al. [25] proposed a EPRY-FPM algorithm to reconstruct both the Fourier spectrum of the sample and the coherent transfer function (CTF), which shares root with ePIE algorithm [8] in conventional ptychography and has been widely used with great performance. In this case, an aberration-free image of the sample can be recovered and spatial-varied aberration of the imaging system can be estimated from the recovered CTF directly. As for LED intensity fluctuation, Bian et al. [26] proposed a LED intensity correction method to solve this problem, which is pretty concise and does not trade with the computational efficiency. Thus, it's easy to be applied without extra computational cost. Note that the problem of slight LED intensity fluctuation won't severely affect the final recovery results during the process of a few minutes of data acquisition. Moreover, the EPRY-FPM can further improve the robustness towards the intensity fluctuation by transforming the error of intensity fluctuation to the error of CTF due to the computational mechanism. But it doesn't work in some specific situations. For example, when a few LEDs attenuate or break down during the process of automatic collection, or setting the different exposure time for bright-field (BF) and dark-field (DF) images captured with 8-bit imaging sensor, all of which will generate remarkable intensity fluctuations in recorded images. In regard to the problem of parameter imperfections, including the error of the distance between adjacent LED elements, the shift of center LED to optic axis, the rotation of LED array and the height between LED array and the sample, Sun et al. [27] proposed a PC-FPM algorithm to correct the positional misalignment, which got perfect performance with the simulated annealing (SA) algorithm and non-linear regression process. Further, the general solutions to FPM [1-3, 25], termed alternating projection (AP) methods, are easy to be implemented and fast to converge, but they are sensitive to measurement noise, which are often attributed to the low signal-to-noise ratio (SNR) of DF images and the non-convex nature of phase retrieval problems [28]. To tackle measurement noise, multiple algorithms employing the difference map [7], nonlinear optimization [29], Gauss-Newton search [30, 31], Wirtinger flows [32], or convex lifting [33] have been proposed to either achieve better convergence properties, improve the robustness towards noise, or both. However, those methods usually rely on expensive processing requirements, making them less appealing from a computational point of view [34]. Therefore, Zuo et al. [34] designed an adaptive step-size strategy based on the AP methods for noise-robust FPM since the AP methods are more favored and widely used for their fast convergence, computational efficiency and low memory cost, especially for dealing with large dataset.

Although various error sources have been researched and some special methods have been proposed respectively, there is no method to solve all of them. It's worth mentioning that any one of those error sources almost leads to similar artifacts [25-27, 31, 34]. Hence it is difficult to distinguish a kind of error source from another only according to the experimental results, which may be degraded by any of aberrations, LED intensity fluctuation, parameter imperfections or the noise. Besides, the systematic error is generally a mixture of various sources in the real situation. As a consequence, the recovery results are less than satisfactory when using either of the mentioned methods to correct mixed error in simulations and experiments. Fortunately, though the AP methods' susceptibility to noise is well-noticed, its offer great flexibilities to be adapted to more complicate mathematical models for many advanced applications [20-22, 34]. Therefore, we purpose a novel system calibration procedure, termed SC-FPM, based on the simulated annealing (SA) algorithm, LED intensity correction and adaptive step-size strategy, which involves the evaluation of an error matric at each iteration step, followed by the re-estimation of accurate parameters to numerically correct the mixed error. It is worth noting that the mixed systematic error can't be addressed by separating each error source because of their inter-relationships and the same artifacts. So all the mixed error sources must be conducted simultaneously. Similar to many proposed correction methods [25-27, 34], the error metric is often used for monitoring the process or the convergence of standard PIE and ePIE algorithms [4, 8], evaluating the performance of correction methods at each iteration step, and updating parameter estimation and the step-size for the next cycle. At first, a number of initial iterations for BF images with low illumination NAs are implemented to correct those low-frequency apertures' parameters and the intensity of raw images accordingly with SA algorithm and LED intensity correction method respectively, which will get more precise initial parameters in that the BF images suffer less from the noise. Next, from the global perspective, re-estimating the parameters to enhance iterative efficiency, as well as adjusting accuracy through the non-linear regression process. Finally, all the captured images are iterated several times for more precise parameters and intensity of recorded images together with the adaptive step-size strategy to resist the fluctuation of final results influenced by the noise. As verified through numerical simulations and experiments, the proposed system calibration procedure improves the robustness of FPM and relaxes the experiment conditions, making the FPM not only more practical and flexible, but also providing new sights to be implemented in various FP imaging platforms.

The remainder of this paper is structured as follows: the principle and the computational mechanism of FPM, as well as the global model of parameter imperfections are described in Section 2.1. Then to further explain our correction procedure, we discuss the conflict between aberration correction and LED intensity fluctuation correction and give our solution to the problem in Section 2.2. Next we introduce our system calibration procedure SC-FPM at length in Section 2.3. After that, the great performance achieved both in simulation and experiments are demonstrated in section 3 and section 4. Finally, conclusions and discussions are summarized in Section 5.

## 2. Methods

*2.1 Principle of FPM*

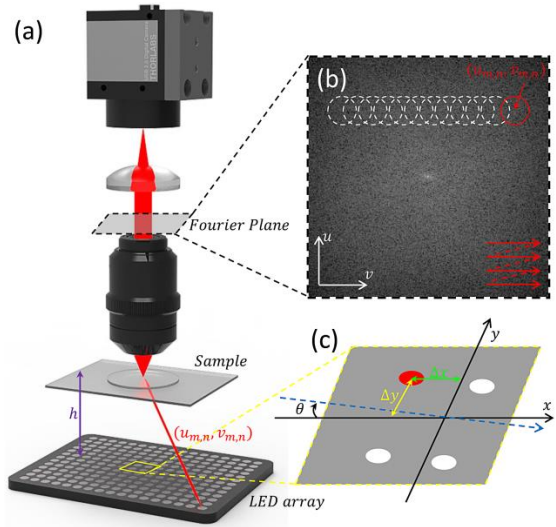

Fig.1. The configuration of a typical FPM setup. (a) An LED array sequentially illuminates the sample with various incident angles. (b) The OTF (red circle) is shifted accordingly from angle-varied illumination in the Fourier plane. The Fourier transform of many shifted LR images (each circle) are iteratively stitched together to extend the complex sample spectrum's resolution beyond the objective lens's cutoff. (c) The model of parameter imperfections with the rotation factor $\theta$, shift factors along x and y axis $\Delta x, \Delta y$ and height factor $h$, where the black 2D coordinates are the ideal coordinates, while the blue dotted line is the real coordinate and the red LED is the central LED in this array.

It's necessary to review the basic principle of FPM reconstruction process in detail in order to explain the mixed systematic error correction method for FP algorithm. A typical FPM setup consists of an LED array, a light microscope with a low-NA objective lens and a monochromatic imaging sensor as shown in Fig.1(a). The LED elements on the array are turned on sequentially to illuminate the sample from various incident angles, which will lead to relative movement between the spectrum of the sample and the aperture of objective lens. Fig.1(b) shows the relative shift of the CTF along with the sequentially lighted LED elements.

Numerically, for each $LED_{m,n}$ (row $m$, column $n$) and its illumination wave vector $(u_{m,n}, v_{m,n})$, under the assumption that the incident light is an ideal plane wave and the sample is relatively thin, the transmitted wave field from the sample $o(x, y)$ can be described as $o(x, y)e^{(jxu_{m,n}, jyv_{m,n})}$, where $o$ is the sample's transmission function, $(x, y)$ are the 2D Cartesian coordinate system in the sample plane, $j$ is the imaginary unit. Then the transmitted light field is Fourier transformed to the Fourier plane and multiplied by OTF accordingly, which can be represented as $\mathcal{F}\{o(x, y)e^{(jxu_{m,n}, jyv_{m,n})}\}P(u,v)$, where $P(u,v)$ is the OTF, which act as a low-pass filter of an imaging system. $(u,v)$ are the 2D spatial frequency coordinates in the Fourier plane, and $\mathcal{F}$ is the Fourier transform operator. Afterward, the light field is inverse Fourier transformed to the image plane and the imaging sensor captures an LR intensity image $I^c_{m,n}$ of the sample, which is given by:

$$I^c_{m,n}(x, y) = \left| \mathcal{F}^{-1}\{O(u-u_{m,n}, v-v_{m,n}) \cdot P(u,v)\} \right|^2 \quad (1)$$

where $\mathcal{F}^{-1}$ is the inverse Fourier transform operator, and $O$ is the Fourier spectrum of the sample's transmission function $o$. The incident wave vector $(u_{m,n}, v_{m,n})$ can be expressed as

$$u_{m,n} = \frac{2\pi}{\lambda} \frac{x_0 - x_{m,n}}{\sqrt{(x_0 - x_{m,n})^2 + (y_0 - y_{m,n})^2 + h^2}}$$
$$v_{m,n} = \frac{2\pi}{\lambda} \frac{y_0 - y_{m,n}}{\sqrt{(x_0 - x_{m,n})^2 + (y_0 - y_{m,n})^2 + h^2}} \quad (2)$$

where $(x_0, y_0)$ is the central position of each small segment, $x_{m,n}$ and $y_{m,n}$ denote the position of the LED element on the row $m$, column $n$. $\lambda$ is the illumination wavelength, $h$ is the distance between the LED array and sample. Note that the Eq.(2) is not uniform in relation to the specific optical system.

Then these LR measurements are iteratively stitched together in the Fourier space using the conventional FP reconstruction algorithm. First, an initial HR complex amplitude guess of the CTF and sample spectrum, labelled as $P_0(u,v)$ and $O_o(u,v)$, are provided to start the algorithm. Generally, the initial guess of CTF is set as a circular shape low-pass filter with all ones inside the pass band, zeros out of the pass band and uniform zero phase. The Fourier transform of a frame of an up-sampled LR image is taken as the initial sample spectrum guess. Second, the exit wave at the Fourier plane can be estimated by the multiplication: $\varphi^e_{m,n}(u,v) = O_0(u-u_{m,n}, v-v_{m,n})P_0(u,v)$, and the simulated LR image on the image plane is the inverse Fourier transform of it: $\phi^e_{m,n}(x,y) = \mathcal{F}^{-1}\{\varphi^e_{m,n}(u,v)\}$. Next, the modulus of the simulated LR image is replaced by the square-root of the real measurement $I^c_{m,n}$ and the phase is kept unchanged, such that:

$$\phi^u_{m,n}(x, y) = \sqrt{I^c_{m,n}(x, y)} \frac{\phi^e_{m,n}(x, y)}{\left|\phi^e_{m,n}(x, y)\right|} \quad (3)$$

The updated LR image $\phi^u_{m,n}(x, y)$ is then used for updating the corresponding spectrum region of the sample estimation as the fourth step, which is given by [21, 27]:

$$O_i(u-u_{m,n}, v-v_{m,n}) = O_i(u-u_{m,n}, v-v_{m,n}) + \alpha \frac{|P_i(u,v)| P_i^*(u,v)}{|P_i(u,v)|_{max}\left(|P_i(u,v)|^2 + \delta_1\right)} \Delta\varphi_{i,m,n} \quad (4)$$

$$P_i(u,v) = P_i(u,v) + \beta \frac{|O_i(u-u_{m,n}, v-v_{m,n})| O_i^*(u-u_{m,n}, v-v_{m,n})}{|O_i(u-u_{m,n}, v-v_{m,n})|_{max} \left(|O_i(u-u_{m,n}, v-v_{m,n})|^2 + \delta_2\right)} \Delta\varphi_{i,m,n} \quad (5)$$

where $\alpha$ and $\beta$ are the step size of the update and usually $\alpha = \beta = 1$ is used for the results [4, 8]. $\delta_1$ and $\delta_2$ are regularization constants to prevent the denominator to be zero. $i$ is the time of iterations. $\Delta\varphi_{i,m,n}$ is the auxiliary function for updating process: $\Delta\varphi_{i,m,n} = \mathcal{F}\{\phi_{i,m,n}^u(x,y)\} - \varphi_{i,m,n}^e(u,v)$.

For single iteration process, repeating above steps until all the captured images $I_{m,n}^c$ are used to update the CTF and sample spectrum. Subsequently, the whole iterative process is repeated for $i$ times until the solution convergences, which is judged by the evaluation of an error matric at each iteration step indicated in Eq.(6). Finally, the sample spectrum is inverse Fourier transformed back to the spatial space, where a HR intensity distribution and phase distribution are recovered.

$$E_i = \frac{\sum_{x,y,m,n} \left(|\phi_{i,m,n}^e(x,y)|^2 - I_{m,n}^c(x,y)\right)^2}{\sum_{x,y,m,n} I_{m,n}^c(x,y)} \quad (6)$$

In the ideal FPM setup, parameters are usually accurate. But in the real situation parameters are misaligned in a variety of forms. Figure 1(c) presents the model of parameter imperfections with the rotation factor $\theta$, shift factors along x- and y-axis $\Delta x, \Delta y$ and height factor $h$ to describe them mathematically. In fact, considering the great performance of pcFPM [27], these four global variables are enough to establish the parameter imperfections model. Additional global or partial variables, such as pitch angle and the distance between adjacent LED elements, certainly can be added to this model due to the same computational mechanism. However, it may increase the computational burden. The black 2D coordinates are the ideal coordinates, while the blue dotted line is the real coordinate and the red LED is the central LED in this array. Then we can express the position of each LED element as

$$\begin{aligned} x_{m,n} &= d_{LED}\left[\cos(\theta)m + \sin(\theta)n\right] + \Delta x \\ y_{m,n} &= d_{LED}\left[-\sin(\theta)m + \cos(\theta)n\right] + \Delta y \end{aligned} \quad (7)$$

where $d_{LED}$ presents the distance between adjacent LED elements. In this paper, we set $d_{LED} = 4mm$ in both simulations and experiments.

### 2.2 Aberrations and LED intensity fluctuation correction

There are two things need to be noticed. Note that the Eq. (4) and (5) are based on the PIE algorithm in our procedure which is quite different from the ePIE-based EPRY-FPM algorithm [25]. In fact, both PIE and ePIE algorithm are widely used but the PIE-based algorithm, namely Eq.(4) and Eq.(5), will be more robust to noise due to the proper evaluation of $\delta_1$ and $\delta_2$ as shown in Fig.2. A programmable $15 \times 15$ LED matrix is used to provide angle-varied illuminations, which means $m \in \{-7,...,0,...,7\}$, $n \in \{-7,...,0,...,7\}$. The parameters in simulations are practically chosen to model a light microscope, with an incident wavelength of $632nm$, an $4\times$ objective with 0.1 NA, an image sensor with $6.5\mu m$ pixel size, and a small segment of $128 \times 128$ pixels. HR input intensity and phase profiles are shown in Fig.2(a) and Fig.2(b). They serve as the ground truth of the simulated complex sample. The distance between the sample and LED array is $86mm$. Besides the idealized situation, each LR image is corrupted with 40% Gaussian noise with different variances to an extreme level. A set of 225 LR intensity images is simulated under this setting. The noise level is quantified by the average mean absolute error (MAE), defined as $AMAE = \langle |I_n - I| \rangle / \langle I \rangle$, where $\langle I \rangle$ is the mean value of all noise-free DF intensity images and $\langle |I_n - I| \rangle$ is the averaged mean absolute error of the corresponding noisy images. The intensity and phase reconstruction accuracy is evaluated by root-mean-square error (RMSE).

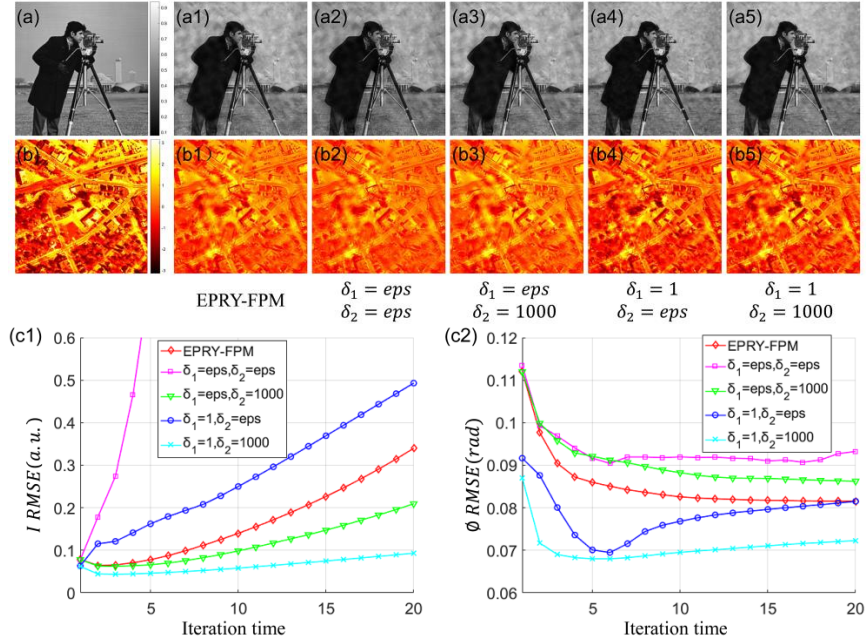

Fig.2. Comparison of intensity (a) and phase (b) recovery results with 40% Gaussian noise using different parameters. (a1) and (b1) are the best results with EPRY-FPM algorithm at 17 iterations. (a2) and (b2) are the best results with $\delta_1 = eps, \delta_2 = eps$ at 6 iterations. (a3) and (b3) are the best results with $\delta_1 = eps, \delta_2 = 1000$ at 20 iterations. (a4) and (b4) are the best results with $\delta_1 = 1, \delta_2 = eps$ at 6 iterations. (a5) and (b5) are the best results with $\delta_1 = 1, \delta_2 = 1000$ at 5 iterations. (c1) and (c2) are the intensity and phase reconstruction accuracy versus iteration time for different algorithms.

The recovery results with 40% Gaussian noise are shown in Fig.2. Obviously, the recovery results can't be converged because of the extreme noise but the best performance will appear at a specific iteration as shown in Fig.2(c1) and Fig.2(c2). Fig.2(a1) and Fig.2(b1) present the best results with ePIE-based EPRY-FPM algorithm at 17 iterations. Fig.2(a2) and Fig.2(b2) present the best results with PIE-based algorithm at 6 iterations where $\delta_1 = eps, \delta_2 = eps$. The epsilon of the machine (eps) is the minimum distance that two numbers could be distinguished by a floating point arithmetic program like Matlab. Similarly, Fig.2(a3) and Fig.2(b3) present the best results with $\delta_1 = eps, \delta_2 = 1000$ at 20 iterations. Fig.2(a4) and Fig.2(b4) present the best results with $\delta_1 = 1, \delta_2 = eps$ at 6 iterations. Fig.2(a5) and Fig.2(b5) present the best results with $\delta_1 = 1, \delta_2 = 1000$ at 5 iterations. Fig.2(c1) and Fig.2(c2) present the intensity and phase reconstruction accuracy versus iteration time for different algorithms. It can be seen that compared with the ePIE-based algorithm, the PIE-based algorithm may be more robust to noise when setting $\delta_1 = 1, \delta_2 = 1000$. Additionally, we test a lot of different combinations of $\delta_1$ and $\delta_2$, but consistently, all these data indicate that the best robustness and convergence efficiency are achieved at $\delta_1 = 1, \delta_2 = 1000$. So we set $\delta_1 = 1, \delta_2 = 1000$ in our following procedure.

Another important issue is the conflict between aberration correction and LED intensity fluctuation correction. If there exists the problem of LED intensity fluctuation, then the Eq.(1) need to be accordingly modified to

$$I_{m,n}^u(x,y) = c_{m,n} \cdot \left| \mathcal{F}^{-1}\{O(u-u_{m,n}, v-v_{m,n}) \cdot P(u,v)\} \right|^2 \tag{8}$$

Where $c_{m,n}$ is defined as [26]

$$c_{m,n} = \frac{\sum_{x,y} |\phi_{m,n}^e(x,y)|^2}{\sum_{x,y} I_{m,n}^c(x,y)} \tag{9}$$

Then the captured intensity images are updated by

$$I_{m,n}^u(x,y) = c_{m,n} \cdot I_{m,n}^c(x,y) \tag{10}$$

The LED intensity correction method is pretty concise and quite pragmatic without sacrificing its running efficiency and computation time. First, calculating the $c_{m,n}$ after the first iteration and update the raw images for the second iteration, then repeating the process until the algorithm converges to a certain degree [26]. Based on the ignored principle of energy conservation, the ratio $c_{m,n}$ ensures that the energy of image plane remain unchanged during the update process as presented in Eq.(3), where the modulus of simulated LR image is replaced by the square-root of the real measurement $I_{m,n}^c(x,y)$. However, a large quantity of simulation data indicates that the original intensity correction algorithm is only valid without the process of updating the CTF. That is, if using the Eq.(5) together with Eq.(4) to update the CTF and sample spectrum simultaneously as shown in Fig.3, the algorithm can't achieve good performance because of the mutual transformation between the error of LED intensity and the error of CTF. So there exists strong conflict between aberration correction and LED intensity correction, which is against our proposed SC-FPM framework. For illustration purpose only, extreme 200% intensity fluctuation is artificially introduced in Fig.3 by multiplying each raw image with a random constant.

Fig.3(a1), (b1) and (c1) present the blurry results of intensity, phase and spectrum respectively without LED intensity correction. We can clearly observe the artifacts of those low-frequency apertures in Fig.3(c1). Fig.3(a2), (b2) and (c2) present the recovery results with the original LED intensity correction method at 30 iterations, which can solve the problem effectively from the subjectively visual perspective and the artifacts in Fig.3(c2) are eliminated. Fig.3(a3), (b3) and (c3) present the recovery results with ePIE-based intensity correction method at first iteration, while Fig.3(a4), (b4) and (c4) show the recovery results at 30 iterations. Fig.3(a5), (b5) and (c5) present the recovery results with PIE-based intensity correction method with $\delta_1 = 1, \delta_2 = 1000$ at first iteration, while Fig.3(a6), (b6) and (c6) present the recovery results at 30 iterations. It is observed that no matter combining LED intensity correction method with the ePIE-based algorithm or PIE-based algorithm, the reconstruction result are less than satisfactory for the reasons mentioned above, and unexpectedly, the best results appear at the first iteration. Fig.3(d1) and Fig.3(d2) present the intensity and phase reconstruction accuracy versus iteration time for different algorithms. The results of ePIE-based algorithm (green line) and PIE-based algorithm (blue line) are quite similar but PIE-based algorithm is more robust, which draws the same conclusion with Fig.2. In addition, note that even with original LED intensity correction method, the recovery results are extremely unstable with sharp oscillation as shown by red line in Fig.3(d1) and Fig.3(d2). In fact, the Eq.(10) is not from the overall perspective, the intensity of each raw image is updated by adjusting their respective coefficient. Note that if multiplying each raw image with the same constant, the constant can be ignored in the FPM model which will have no effect on the final reconstruction. But if making the original LED intensity correction method more robust and also valid for aberration correction, it needs to be modified with a global perspective. First it directly comes to mind that the updating operation needs to run at the same standard, which is introduced by setting the reference intensity. In addition, for reducing the conflict between aberration correction and intensity correction, the PIE-based intensity correction algorithm will be implemented only once. So the modified solution is as follows.

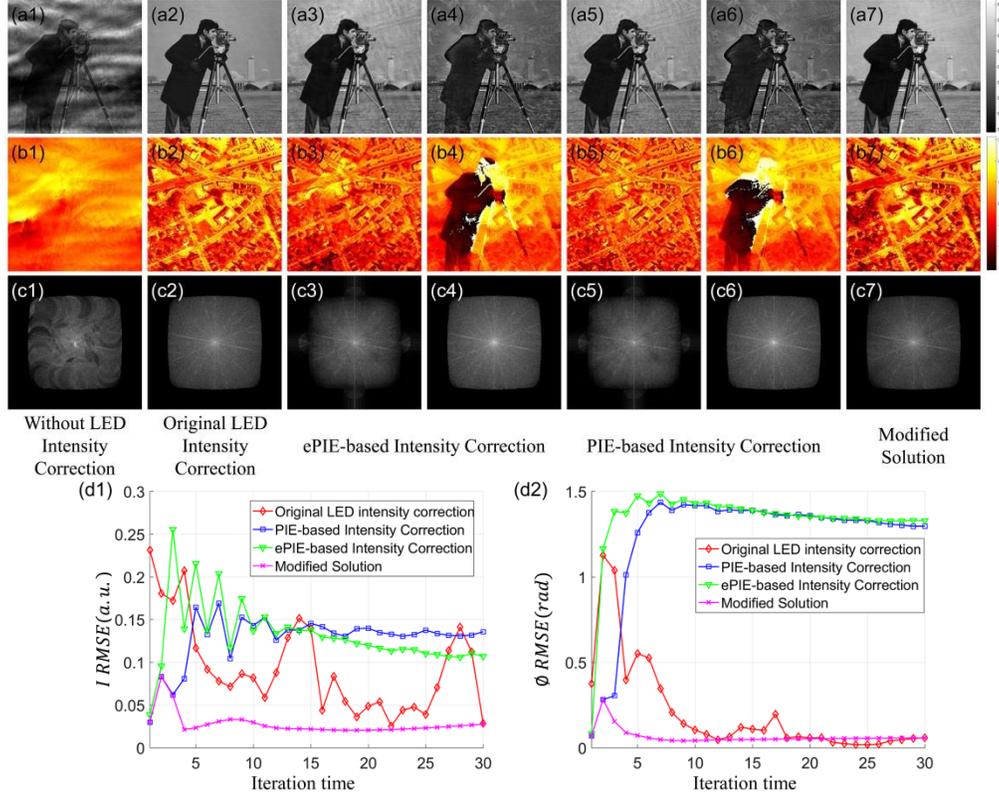

Fig.3. Conflict and solution between aberration correction and LED intensity correction. (a1), (b1) and (c1) present the recovery results of intensity, phase and spectrum respectively under 200% LED intensity fluctuation to an extreme degree. (a2), (b2) and (c2) present the recovery results with LED intensity correction method only with Eq.(4) at 30 iterations. (a3), (b3) and (c3) present the recovery results with LED intensity correction using EPRY-FPM algorithm at first iteration, while Fig.3(a4), (b4) and (c4) show the recovery results at 30 iterations. Fig.3(a5), (b5) and (c5) present the recovery results with LED intensity correction with $\delta_1 = 1, \delta_2 = 1000$ at first iteration, while Fig.3(a6), (b6) and (c6) present the recovery results at 30 iterations. (a7), (b7) and (c7) present the recovery results with our solution at 9 iterations. (d1) and (d2) present the intensity and phase reconstruction accuracy versus iteration time for different algorithms.

First, calculate the ratio $c_{m,n}$ after the first iteration and update the raw images for the second iteration. After that, without intensity correction process, calibrating the aberration only through Eq.(4) and Eq.(5) for the rest iterations. Here the center LED is set as the reference which is supposed to be free of intensity fluctuation. Then the Eq.(9) needs to be modified to

$$c_{m,n} = \frac{\sum_{x,y} \left| \phi^e_{m,n}(x,y) \right|^2}{c_{0,0} \cdot \sum_{x,y} I^c_{m,n}(x,y)} \qquad (m,n \neq 0) \qquad (11)$$

Where $c_{0,0} = \sum_{x,y}|\phi_{0,0}^e(x,y)|^2 / \sum_{x,y} I_{0,0}^c(x,y)$ and the update operation is

$$I_{m,n}^u(x,y) = \begin{cases} c_{m,n} \cdot I_{m,n}^c(x,y) & (m,n \neq 0) \\ I_{0,0}^c(x,y) & (m = n = 0) \end{cases} \quad (12)$$

Accordingly, the best final reconstruction with our modified LED intensity correction method are shown as Fig.3(a7), (b7) and (c7) at 9 iterations. Compared with Fig.3(a2), (b2) and (c2), the results (pink line) are more stable, accurate and efficient than the original LED intensity correction method (red line) as the result of Fig.3(d1) and Fig.3(d2).

*2.3 System calibration algorithm framework (SC-FPM)*

Sever artifacts will be aroused in reconstruction results due to systematic error, which mainly comes from aberrations, LED intensity fluctuation, parameter imperfections and the noise. Although these problems have been researched and some special methods have been proposed respectively, there is no method to solve all of them. Referring to these specific methods respectively [25-27, 34], here we proposed the SC-FPM algorithm to solve the mixed four errors. The overall framework is PIE-based algorithm introduced in Section 2.1, which is including the process of updating the CTF and more robust compared with ePIE-based algorithm proved in Section 2.2. Then we add the LED intensity correction method as described in details at the end of Section 2.2. So far, the problems of aberration and LED intensity fluctuation have been addressed simultaneously. In this section, we'll mainly focus on the calibration of parameter imperfections and the noise. The model of parameter imperfections has been discussed at the end of Section 2.1. Figure 4 shows a flow chart of SC-FPM procedure.

At first, similar to the typical FPM algorithm depicted in Section 2.1, an initial guess of the sample spectrum $O_0(u,v)$ and CTF function $P_0(u,v)$ are provided to start the algorithm. Second, we define the LED updating range $S_i$ for each iteration. Normally, the updating process should repeat for all the 225 images until each incident angle has been addressed, which is regarded as once iteration of the algorithm and processed in turn with both the sample spectrum and CTF updated in each loop. However, the low-frequency information is more important to the reconstruction which also decides the processing order of the captured images $I_{m,n}^c(x,y)$. So a number of initial iterations for BF images with low illumination NAs are implemented to correct those low-frequency apertures' parameters, as well as update those raw images according to modified intensity correction method proposed in Section 2.2. More precise parameters can be achieved by implementing BF images since that they suffer less from the noise. So it'll be effective to separate the noise and parameter imperfections. For SC-FPM, in the first ten iterations, where $i=1,\ldots,10$, the process repeats for $5 \times 5$ BF images with the LED updating range $S_i = \{(m,n)|m=-2,...,2,n=-2,...,2\}$. Since only 25 images are computed in each iteration, the initial value of the four global factors $(\theta, \Delta x, \Delta y, h)$ can be efficiently obtained within ten iterations. It should be noted that different choice of parameters in SC-FPM would affect its correction accuracy and efficiency. Generally, the apertures' positions of those 25 images could be accurately corrected after ten initial iterations under the extreme conditions in our simulation. Thus we implement ten initial iterations in this manuscript empirically. At the end of each initial iteration, $O_i(u,v)$ and $P_i(u,v)$ need to be initialized because the object's profile could be extremely distorted when those 25 apertures' positions have not been corrected properly. Besides, $I_{m,n}^c(x,y)$ also needs to be initialized due to the same reason but according to the LED intensity correction method, the first iteration is used to calculate the ratio $c_{m,n}$ and update the raw images for next iteration.

So we take two iterations as one group and initialize the $I^c_{m,n}(x,y)$ at the end of each even iteration before next updating. After ten initial iterations, all the 225 recorded images are iterated for precise position correction with the adaptive step-size strategy to resist the fluctuation of final results influenced by noise. And the LED intensity correction method will be only used for once with all the 225 images without initialization at the twelfth iteration. Therefore, in SC-FPM, the LED updating range $S_i$ for each iteration is defined as

$$S_i = \begin{cases} \{(m,n)|m=-2,...,2, n=-2,...,2\} & i \leq 10 \\ \{(m,n)|m=-7,...,7, n=-7,...,7\} & else \end{cases} \quad (13)$$

Next, Simulated Annealing (SA) algorithm is used to adjust the incident angle of $LED_{m,n}$ in the Fourier plane to calibrate parameter imperfections. Firstly, a set of further estimates of the frequency aperture $\varphi^e_{r,i,m,n}(u,v)$ are calculated, each with a random frequency-offsets $(\Delta u_{r,m,n}, \Delta v_{r,m,n})$, where $r=1,...,8$. Here $r$ represents eight different frequency-shifting directions, and each $(\Delta u_{r,m,n}, \Delta v_{r,m,n})$ is a random frequency-shifting distance between $\pm \Delta_{i,u,v}$ along the $r$th direction. Afterwards, the $r$th estimate of frequency aperture is computed as

$$\varphi^e_{r,i,m,n}(u,v) = O_i\left(u-(u_{m,n}+\Delta u_{r,m,n}), v-(v_{m,n}+\Delta v_{r,m,n})\right) P_i(u,v) \quad (14)$$

and the simulated image on the imaging sensor is $\phi^e_{r,i,m,n}(x,y) = \mathcal{F}^{-1}\{\varphi^e_{r,i,m,n}(u,v)\}$. Then the intensity distribution of each $\phi^e_{r,i,m,n}(x,y)$ is compared with $I^c_{m,n}(x,y)$ to give a set of error matric:

$$E(r) = \sum_{x,y}\left(|\phi^e_{r,i,m,n}(x,y)|^2 - I^c_{m,n}(x,y)\right)^2 \quad (15)$$

The intensity distribution of the simulated image should be approximate to the captured image, so smaller value of $E(r)$ stands for better recovery performance. Here the index of the minimum value of $E(r)$ is labeled as $s$ and the frequency aperture's position can be updated as

$$\begin{aligned} s &= \arg\min[E(r)] \\ u^u_{m,n} &= u_{m,n} + \Delta u_{s,m,n} \\ v^u_{m,n} &= v_{m,n} + \Delta v_{s,m,n} \end{aligned} \quad (16)$$

The variable $\pm\Delta_{i,u,v}$ begins at a predefined value and will gradually decrease to a small (or zero) value with a set number of iterations, which is defined as the searching step length of SA algorithm. And we choose $\Delta_{1,u,v} = 8$ in our procedure due to the introduction of the extreme systematic error. Then the step length is updated as follows. In the first ten initialized iterations (five groups), decreasing it by half to compress the frequency searching range at each odd iterations. In the meanwhile, it is supposed to be no less than two before the global 225 images are iterated.

$$\Delta_{i+1,u,v} = \begin{cases} \dfrac{\Delta_{i,u,v}}{2} & i=3,5 \\ \Delta_{i,u,v} & i=2,4,6 \\ 2 & i=7,...,10 \\ 1 & else \end{cases} \quad (17)$$

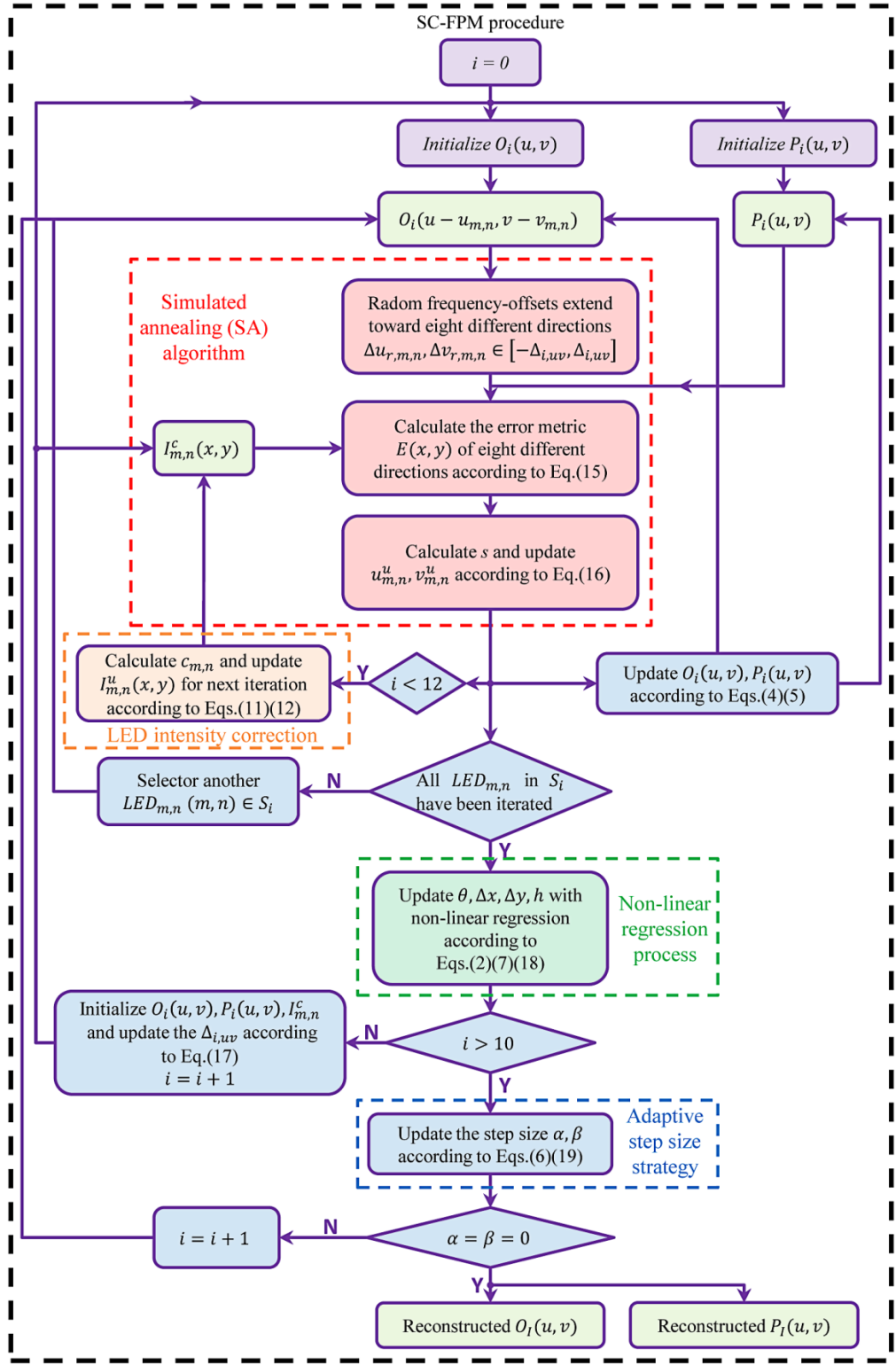

Fig.4. Flow chart of SC-FPM method

Consequently, from the global perspective, the parameters are re-estimated to achieve the enhancements of the iterating efficiency and adjusting accuracy with the non-linear regression [27] which can be expressed as

$$Q(\theta, \Delta x, \Delta y, h) = \sum_{m,n} \left[ \left( u_{m,n}(\theta, \Delta x, \Delta y, h) - u_{m,n}^u \right)^2 + \left( v_{m,n}(\theta, \Delta x, \Delta y, h) - v_{m,n}^u \right)^2 \right]$$
$$(\theta, \Delta x, \Delta y, h)^u = \arg min \left[ Q(\theta, \Delta x, \Delta y, h) \right] \tag{18}$$

where $Q(\theta, \Delta x, \Delta y, h)$ is the defined non-linear regression function which needs to be minimized. $\left[ u_{m,n}(\theta, \Delta x, \Delta y, h), v_{m,n}(\theta, \Delta x, \Delta y, h) \right]$ denotes the updated position of the frequency aperture corresponding to the illumination LED element on row *m* column *n*, which is very sensitive to the misaligned parameters. And $(\theta, \Delta x, \Delta y, h)^u$ are the updated global positional factors through the non-linear regression process. Finally, the adaptive step-size strategy is used to resist the fluctuation of final results resulting from the noise by decreasing $\alpha, \beta$, the step length of the algorithm in Eq.(4) and Eq.(5). The update operation is as follows according to Eq.(6). Similarly, $\beta$ is updated in the same way.

$$\alpha_i = \begin{cases} \alpha_i & (E_i - E_{i-1})/E_{i-1} > 0.1 \\ \dfrac{\alpha_i}{2} & otherwise \end{cases} \tag{19}$$

## 3. Simulations

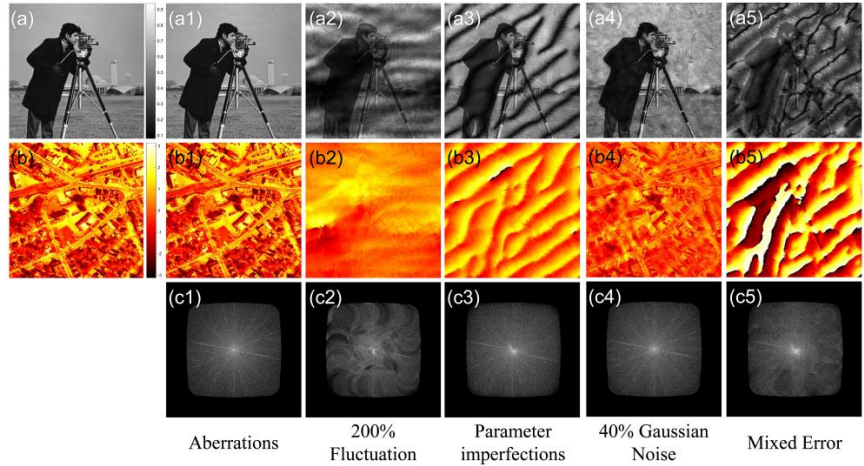

Fig.5. The performance of the original PIE-based algorithm with different systematic error sources. HR input intensity (a) and phase profiles (b) serve as the ground truth of the simulated complex sample. Group (a1), (b1) and (c1) present recovery results with aberrations. Group (a2), (b2) and (c2) present recovery results with 200% LED intensity fluctuation. Group (a3), (b3) and (c3) present recovery results with unknown parameter imperfections. Group (a4), (b4) and (c4) present recovery results with 40% Gaussian noise. Group (a5), (b5) and (c5) present recovery results with mixed four errors.

Before applying SC-FPM to the experimental data, we first evaluate its effectiveness through simulations. The simulation parameters are the same as Section 2.2. For illustrative purposes only, the systematic errors, such as aberration, LED intensity fluctuation, parameter imperfections and the noise are deliberately exaggerated in our simulations. Thus it'll be very intuitive to evaluate the performance and robustness of SC-FPM compared with those specific advanced methods respectively. Generally, the aberrations in an optical system can be decomposed into a set of Zernike polynomials, each with a different coefficient. So the aberrations are introduced by replacing the phase of ideal CTF with a specific Zernike polynomial like $Z_5^1$. In addition, intensity uncertainly is artificially introduced by multiplying each raw image with a random constant. Here 200% intensity fluctuation is introduced by changing the random constant from zero to two. Positional misalignment is artificially introduced by setting the four positional factors with random values. Here we set $\theta = 5°, \Delta x = 1mm, \Delta y = 1mm, h = 87mm$ as the real situation, while $\theta = 0, \Delta x = 0, \Delta y = 0, h = 86mm$ is the ideal condition. And the noise is artificially introduced by corrupting each low resolution image with 40% Gaussian noise with different variances.

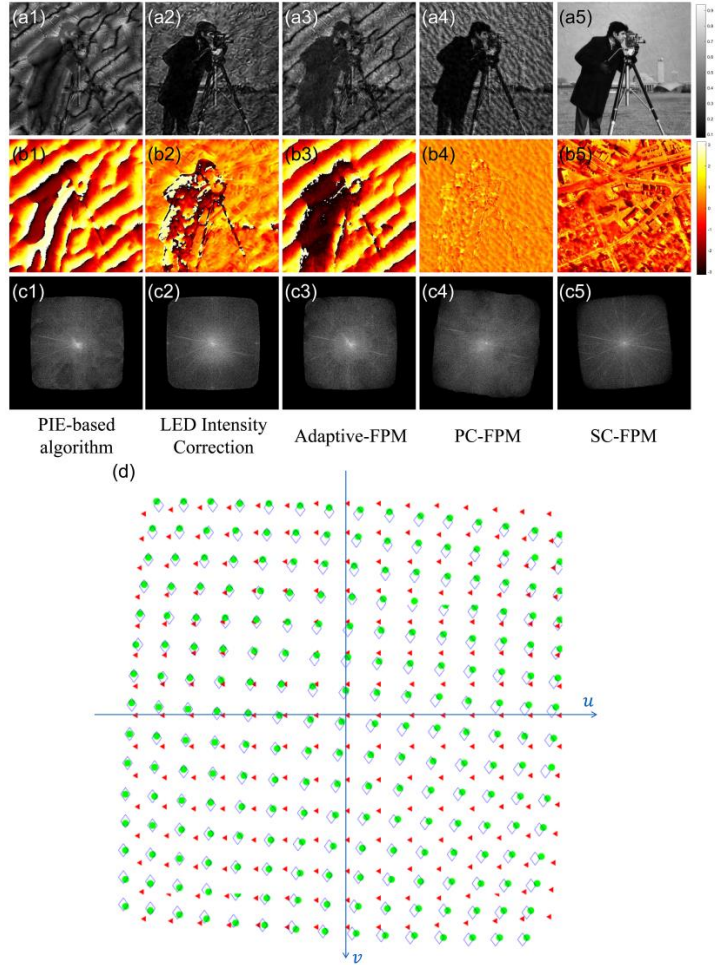

Fig.6. The recovery results of different specific algorithms with mixed systematic error. Group (a1), (b1) and (c1) present recovery results of original PIE-based algorithm. Group (a2), (b2) and (c2) present recovery results of original LED intensity correction method. Group (a3), (b3) and (c3) present recovery results with adaptive step-size FPM. Group (a4), (b4) and (c4) present recovery results with PC-FPM. Group (a5), (b5) and (c5) present recovery results with SC-FPM. (d) presents the position of each frequency aperture corresponding to a specific LED element, where red triangle denotes the ideal model, green point represents the real position and the blue diamond shows the recovery position.

Figure 5 shows the performance of the original PIE-based algorithm with different systematic error sources, each with six iterations. Fig.5(a1), (b1) and (c1) present recovery results only with aberrations. Reasonably, the original PIE-based algorithm has the capability to compensate the aberrations due to its computational mechanism, which updates the sample spectrum and the CTF simultaneously by implementing the Eq.(4) and Eq.(5). As a result, the reconstruction results are free of artifacts and quite approximate to the ground truth as shown in Fig.5(a) and Fig.5(b). However, it's unexpected that other three error sources cause similar artifacts when the systematic errors are not so serious. But in the extreme situations, there are some particular features of each error respectively. Fig.5(a2), (b2) and (c2) present recovery results with 200% LED intensity fluctuation. Compared Fig.5(c1), the recovered sample spectrum has been severely blurred and full of patches, which are the symbol of artifacts caused by LED intensity fluctuation. Fig.5(a3), (b3) and (c3) present recovery results with parameter imperfections where obvious wrinkles can be observed in Fig.5(a3) and (b3), which is quite different from the blurry in Fig.5(a2) and (b2). Besides, the center bright spot of the sample spectrum is somewhat distorted to the top left as shown in Fig.5 (c3). Fig.5(a4), (b4) and (c4) present recovery results with 40% Gaussian noise, which makes the results more even. Synthetically, Fig.5(a5), (b5) and (c5) present recovery results with mixed four errors. All Systematic errors are fused together whether in the spatial domain or frequency domain, each retaining their own features.

Fig.6 shows the performance of different specific algorithms with mixed systematic errors. Undoubtedly all these specific advanced algorithms can't achieve satisfactory performance under such extreme conditions. Fig.6(a1), (b1) and (c1) present recovery results of the original PIE-based algorithm. Fig.6(a2), (b2) and (c2) present recovery results of the original LED intensity correction method, the performance has been enhanced very limitedly compared with Fig.6(a1), (b1) and (c1). Fig.6(a3), (b3) and (c3) present recovery results of adaptive step-size FPM to suppress the noise, but the results are still not satisfactory. Fig.6(a4), (b4) and (c4) present recovery results of PC-FPM to correct parameter imperfections, however, it fails to retrieval the complex sample either and even run into an opposite direction as shown in Fig.6(c4). In fact, although SC-FPM is quite similar to PC-FPM in the part of SA algorithm and non-linear regression process, SC-FPM will be more effective and robust as shown in Fig.6(a5) and (b5) under extreme conditions. Fig.6(d) presents the position of each frequency aperture corresponding to a specific LED element, where red triangle denotes the ideal model, green point represents the real position and the blue diamond shows the recovery position.

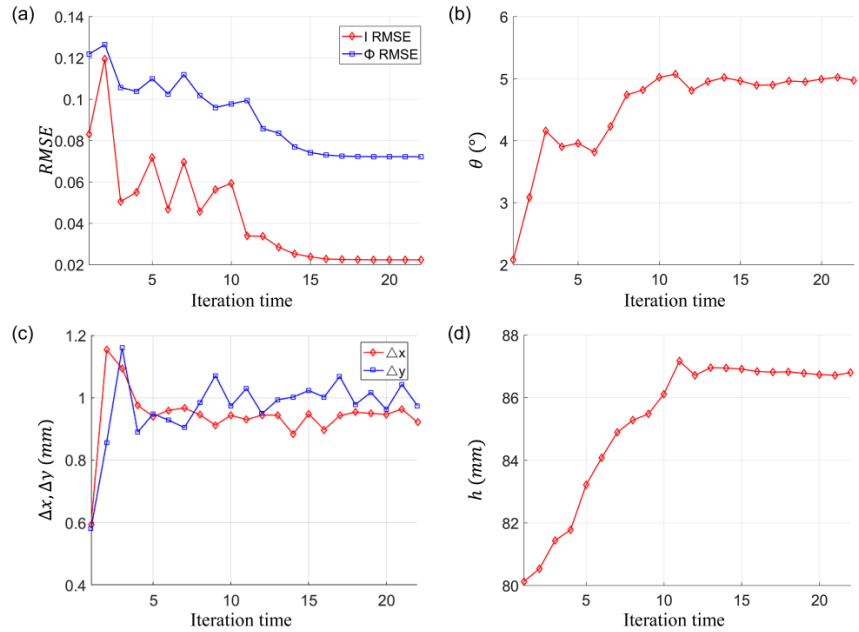

Fig.7. The results of each iteration of SC-FPM in detail. (a) presents the RMSE of intensity and phase versus the iteration time. (b) presents the recovery results of rotation factor $\theta$

versus the iteration time. (c) presents the recovery results of shift factors $\Delta x, \Delta y$ versus iteration time. (d) presents the recovery results of height factor *h* versus iteration time.

Figure 7 shows the detailed results of each iteration in SC-FPM. Within 10 iterations the recovery results are not so stable with BF images as shown in Fig.7(a). After 15 iterations the results tend to be stable. The rotation factor $\theta$ tends to converge on $5°$ around shown in Fig.7(b). The shift factors $\Delta x, \Delta y$ tend to converge on *1mm* shown in Fig.7(c). The height factor *h* tends to converge on *87mm* shown in Fig.7(d). And the final results accurately converge on $\theta = 4.97°, \Delta x = 0.9215mm, \Delta y = 0.9743mm, h = 86.799mm$, which are pretty close to the real parameters introduced at the beginning, proved that the SC-FPM is very robust and well adapted to the real situation.

## 4. Experiments

In order to evaluate the effectiveness of SC-FPM experimentally, we first compare the recovered intensity and phase distribution of one segment ($90 \times 90$ pixels) in a USAF target with different algorithms as shown in Fig.8. All LR images are captured with an $4 \times 0.1$NA objective and a CCD camera ($3.75 \mu m$, DMK23G445, Imaging Source). A programmable $32 \times 32$ LED array (Adafruit, 4mm spacing, controlled by an Arduino) is placed at 86mm beneath the sample. The central 15×15 red (central wavelength 631.13nm with 20nm bandwidth) LEDs are used to provide angle-varied illuminations, resulting in a final synthetic NA of 0.5 theoretically. In simulations, the systematic errors are deliberately magnified in order to verify the effectiveness and robustness of our proposed SC-FPM framework. However, the errors are less obvious in the real situation due to the elaborately experimental operation. On the one hand, it is not likely to correct such small errors by manually adjusting the relative position of each component, unless the system is equipped with a 3D high precision translation stage, which will make the FPM much more costly. And some errors, for instance, the noise are inevitable. On the other hand, insignificant errors will lead to the similar artifacts, making it's difficult to distinguish a kind of error from another only according to the captured images, so the satisfactory performance can't be achieved by applying any of a specific algorithm. Considering this situation, our proposed SC-FPM will be more suitable for dealing with the mixed systematic errors in various forms.

Fig.8(a) presents the FOV of a USAF target captured by an $4 \times 0.1$NA objective, and Fig.8 (a1) shows the enlargement of a sub-region of (a), which becomes blurry since restricted by the low NA of objective. Group (b), (c) and (d) show the recovery results of intensity, phase and spectrum respectively with different algorithms. Fig.8(b1), (c1) and (d1) presents the recovery results of original PIE-based algorithm at 6 iterations, we can conclude that there absolutely exists the parameter imperfections, which is symbolized by obvious wrinkles in the recovered intensity and phase images. While the intensity correction method has failed to reconstruct the sample as shown in Fig.8(b2), (c2) and (d2) at 6 iterations. Fig.8(b3), (c3) and (d3) presents the recovery results of adaptive FPM at 18 iterations, where the intensity image has better performance than original PIE-based algorithm. Furthermore, compared with Fig.8(b2), (c2) and (d2), we can infer that the systematic errors is mainly from the noise and parameter imperfections rather than LED intensity fluctuation. Fig.8(b4), (c4) and (d4) presents the recovery results of PC-FPM at 12 iterations where the obvious wrinkles have been eliminated with the final recovered parameters of $\theta = -1.8°, \Delta x = 1.274mm,$ $\Delta y = 1.270mm, h = 94.399mm$. But the intensity image is less clear than Fig.8(b3), where the group 8 element 4,5,6 can be clearly resolved. Fig.8(b5), (c5) and (d5) presents the recovery results of SC-FPM at 28 iterations with the final recovered parameters of $\theta = -1.6°,$ $\Delta x = -1.188mm, \Delta y = 1.003mm, h = 89.733mm$. Compared with other algorithms, the performance of SC-FPM are more effective and robust to the unknown systematic errors, and the recovered intensity and phase images has higher contrast and resolution where each line pair can be clearly resolved with a uniform background.

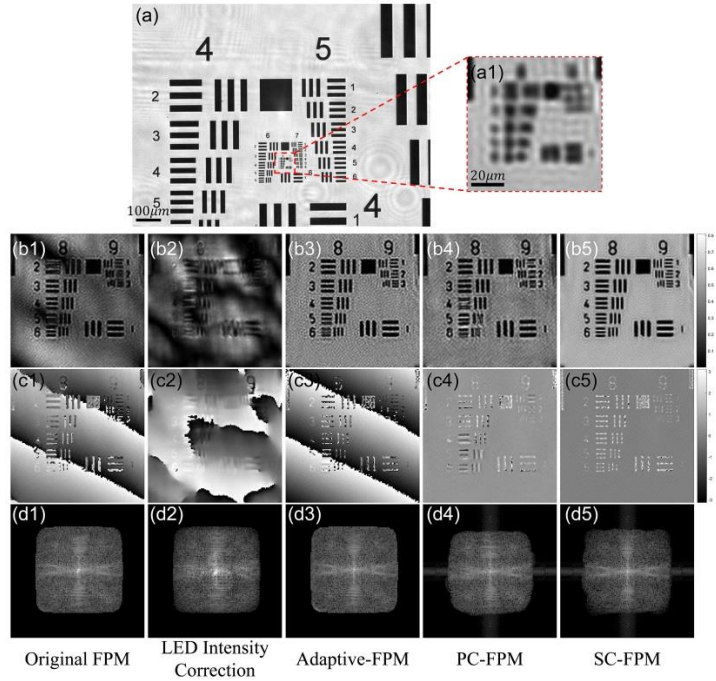

Fig.8. Experimental results of one segments ( $90 \times 90$ pixels) in a USAF target recovered with different advanced algorithms. (a) presents the FOV captured by an $4 \times 0.1$NA objective. (a1) shows the enlargement of a sub-region of (a) as the LR intensity images. Group (b), (c) and (d) show the recovery results of intensity, phase and spectrum respectively with different algorithms.

In addition, we also test our method in in biological sample (Stem transection of dicotyledon) with different algorithms as shown in Fig.9. The LED array is placed at 85.90mm beneath the sample with central red $9 \times 9$ LEDs providing the angle-varied illuminations, resulting in a final synthetic NA of 0.35 theoretically. But the results are quite different from above due to the different systematic errors. Fig.9(b1), (c1) and (d1) presents the recovery results of original PIE-based algorithm at 6 iterations, appearing obvious noise in the recovered intensity and phase images. Fig.9(b2), (c2) and (d2) presents the terrible recovery results of intensity correction method at 6 iterations. Fig.9(b3), (c3) and (d3) presents the recovery results of adaptive FPM at 29 iterations, where the recovered intensity image is dramatically improved compared with Fig.9(b1) but the phase image remains unchanged. Fig.9(b4), (c4) and (d4) presents the recovery results of PC-FPM at 12 iterations with the final recovered parameters of $\theta = -4.9°, \Delta x = -0.104mm, \Delta y = -0.384mm, h = 84.522mm$, where the recovered phase image is quite better than Fig.9(c1) and (c3) but the contrast of the intensity image is very low. Fig.9(b5), (c5) and (d5) presents the recovery results of SC-FPM at 34 iterations with the final recovered parameters of $\theta = -3.4°, \Delta x = -0.427mm, \Delta y = -0.287mm, h = 84.733mm$, where the recovered intensity and phase images show higher contrast and resolution. Fig.8 and Fig.9 together demonstrate the effectiveness and robustness of SC-FPM.

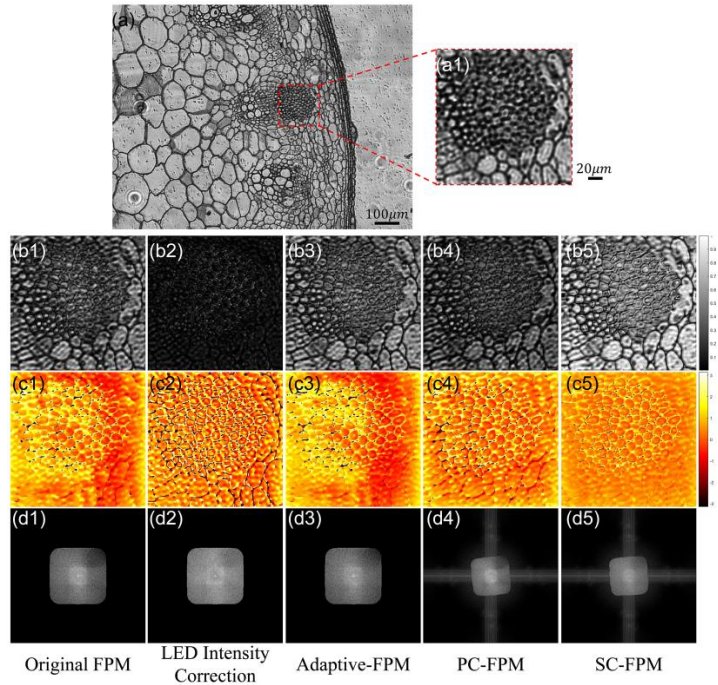

Fig.9. Experimental results of one segments ($200\times 200$ pixels) in a dicotyledonous tissue recovered with different advanced algorithms. (a) presents the FOV captured by an $4\times 0.1$NA objective. (a1) shows the enlargement of a sub-region of (a) as the LR intensity images. Group (b), (c) and (d) show the recovery results of intensity, phase and spectrum respectively with different algorithms.

## 5. Conclusions and discussions

In this paper, we report a system calibration procedure, termed SC-FPM, based on the simulated annealing (SA) algorithm, LED intensity correction method and adaptive step-size strategy theoretically and experimentally. Through combining advantages of each advanced algorithm compatibly, SC-FPM can retrieval a high-quality, noise-robust complex object in the case of extreme multiple errors, including aberrations, LED intensity fluctuation, parameter imperfections and noise in a variety of forms. The effectiveness and robustness of SC-FPM has been verified by simulations and experiments respectively with great performance. Considering that the SC-FPM has many parameters such as the initial iteration time and initial step length of SA algorithm need to set according to real situations, reasonable and effective evaluation of the factors will be conducive to reconstruct with accuracy and efficiency. Besides, we mainly focus on the quality of reconstructions. And the aberration is introduced with a simple Zernike polynomial $Z_5^1$. In fact, the calibration of aberration is another noteworthy issue especially under such mixed system errors, which may be the subject of future work.

## Funding

National Natural Science Fund of China (NSFC) (61377008, 81427802).

## Acknowledgments

The authors thank Jiasong Sun and Chao Zuo for the valuable and helpful discussions and comments.